\documentclass[amsmath,amssymb]{revtex4}
\usepackage[dvips]{graphicx}
\usepackage{epsfig}
\usepackage{subfigure}
\usepackage{graphicx,mathtools,amsmath,amsfonts,amssymb}

\newcommand{\be}{\begin{equation}}
\newcommand{\ee}{\end{equation}}
\newcommand{\bea}{\begin{eqnarray}}
\newcommand{\eea}{\end{eqnarray}}
\usepackage[table]{xcolor}

\begin{document}\
\title{Binding energies  of composite boson clusters  using the Szilard engine}

\author{A. Thilagam}
\affiliation{Information Technology, Engineering and Environment, 
\\
Mawson Institute,\\
University of South Australia, \\
Australia  5095}

\begin{abstract}
We evaluate the binding energies  of systems of bosonic and fermionic particles
on the basis of the  quantum Szilard engine, which confers an energetic value to information and
entropy changes.  We extend  treatment of  the quantum information thermodynamic operation
of the Szilard engine to its  non-trivial role in Bose-Einstein condensation of
 the light mass   polariton quasiparticle, and  binding of large multi-excitonic complexes, 
 and note the same order of magnitudes of exchange and extraction energies 
in these disparate systems. 
We examine the   gradual decline of a defined information capacitive energy
with size of the boson cluster as well as the influence of
confinement effects in composite boson systems.
Can quantum informational entropy changes partly
explain the observations of polariton condensates? We
provide energy estimates using the  system of polariton condensates 
placed in a hypothetical  quantum Szilard engine, and 
discuss the  importance of incorporating 
entropy changes introduced  during quantum measurements, and 
in the interpretation of experimental results.
\end{abstract}

\maketitle

\section{Introduction}

Recently, studies of deviations  from perfect bosonic behavior in composite bosons
or ``cobosons'' \cite{Combe1,Combe2,Combe3}, and from which fermionic  features emerge by virtue of   the Pauli exclusion principle  has received increased attention \cite{Law,Woot,Rama,Tichy1,Tichy2,bobby,thilamc}, partly due to 
the quantum correlated electronics of such systems. While a system of ideal bosons
possess enhanced entanglement features as quantified by the Schmidt number measure, 
 the degree of  correlations  between the particles diminish with
appearance of fermionic attributes. There is 
loss in information that occurs when composite bosons are 
formed, as  illustrated by the inaccessibility 
of spin configuration variables, which otherwise is
explicitly known in the constituent fermions.
There are subtleties  to the degree of loss of information which occurs, for instance,
in a system of two or more bosons: two fermions
constituting a single boson may be distinguished
if the separate bosons occupy distinct external states (position or momentum
of center of mass wise), and entanglement involves pairs of 
different fermions \cite{Rama}. The intrinsically quantum feature of 
distinguishability therefore underpins quantum processing attributes of 
composite bosons.

An area that has been gathering great interest in recent years, and of relevance to
composite boson systems, are the
 links between thermodynamic principles and quantum information entities
 \cite{laud,Szilard,bied,Wook,Yao,Saga1,Saga2,ved1,ved2}. Such links
were first  examined on the basis of  computation energy cost by Landauer \cite{laud}.
Landauer considered that the  irreversible manipulation of information 
is accompanied by a corresponding increase in the tangible
degrees of freedom of system, or the environmental reservoir at the receiving
end of the coded information.  This scheme establishes a vital link
between information exchanges and a  physical mechanism that may take a specific form.
In an earlier work, Szilard employed the  Maxwell's demon model to show that $k_B T\ln 2$ of work 
can be extracted from a thermodynamic cycle, and highlighted that a positive entropy 
production in measurement compensates for the work gain \cite{Szilard}. This ensures that 
the second law of thermodynamics is left intact, by virtue of being
governed by statistical uncertainties.

The hypothetical Szilard engine (shown below in Fig.\ref{eng}) consists of
an atom confined in  box, with the  thermodynamic engine cycle involving the main steps of:
(1) Insertion of a partition to divide the engine box into two disconnected parts, 
(2) Uncertainty in actual position of the atom is acknowledged, 
(3) Measurement procedure to  determine the exact location of the atom, (4) Extraction of  work ($k_BT \ln 2$)
via isothermal expansion by  moving the partition quasi-statically
while in contact with a thermal reservoir of temperature $T$,
and (5) Removal of the partition completing  the engine cycle. 
While the classical engine ignores entropy changes due to
measurement procedures,  the entropy that is  created by the process of observation \cite{bied}
forms a key element of the quantum version of the Szilard engine, 
 highlighting  the critical role of monitoring instruments in the quantum regime.

This work aims to  examine composite boson systems
in the context of  the energy-information link, and 
the application of energetic value of information 
to the  evaluation of the binding energies of exotic systems such
as boson and fermion condensates,
which appears not to have been examined in greater details
in earlier works. Excitons which are
correlated electron-hole quasi-particles possess high binding energies when
confined in semiconductor two-dimensional quantum wells \cite{chem,ding,masu,taka,jai,oh}
where motion is quantized in the direction perpendicular
to the well. A strong-coupling interaction occurs  between excitons and photons
when the quantum well is positioned to coupled with the optical modes of  microcavities,
giving rise to polaritons with integer-spins  \cite{weis,yama,butov}. Polaritons are composed 
of interacting exciton-photon quasi-particles,
which provide an ideal example of a system capable of forming  composite boson condensates.
Due to the photonic contribution (with  polariton resonances around 800 nm in one work \cite{amo}),
the polaritons possess greatly reduced effective masses of 
about 10$^{-5} m_0$ where $m_0$ is the free-electron mass. The
 small mass results in large Broglie wavelengths which exceed the mean average separation of the bosonic polaritons,
 a condition needed for 
Bose-Einstein condensation. Experimental observations have been reported
at lattice temperatures as high as a  few kelvins in several works \cite{lag,bali}, these temperatures in general
are higher than those obtainable for the pure exciton or cold atoms \cite{bosecom,moska,davy}.
It has been shown \cite{lag} that 
condensates can preserve spatial coherences over distances much larger
than the polariton De Broglie wavelength \cite{timo}, hence quantum properties are remarkably revealed 
on classical scales of time and length below a certain transition temperature. 
As a consequence, the condensates
 exhibit superfluidity \cite{amo} where quantum fluid appear to move without great friction,
and composite bosons exhibit superconducting properties \cite{carbo}.
The phase transitions that occur in Bose-Einstein condensates are yet to be 
examined from the context of the deep links between energy and information.
In this work, we use both these experimental values (mass, temperature)
to make comparison of energies of boson systems derived via the Szilard engine.

We point out other excitonic systems, that can be 
examined from the Szilard engine perspective.
For instance, it is well known that large exciton complexes ($E_{N,X}$), with 
$N$ number of neutral excitons bound to an electron 
are known to exist in two-dimensional systems of fermions (electrons,
 holes) in the presence of strong magnetic fields \cite{2d0,2d1,2d2,2d3}.
Such complexes may be present in a
multi-component plasma systems containing both electrons
and $E_{N,X}$ complexes. The appearance of these large complexes, 
may partly be  explained by a model
of localized system of distinguishable particles with negligible wavefunction overlap,
that interact more significantly with  an external  field than with each other.
Under these conditions, the $N$-body problem  may be reduced 
 to an equivalent two-body problem \cite{hall,silve}. In similar light, the
coalescing of the properties of a system of two excitons 
into one exciton has produced good agreement with experimental
results for the singly charged exciton \cite{thilex}
 and biexciton \cite{jai1,jai2}.

In the context of the Szilard engine, it would be worthwhile
to seek clarification on whether the energies of large Coulombic systems such as 
multi-excitonic systems are partly due to quantum informational entropy exchanges.
The dependence of binding energies on  factors such
as the number of charge carriers, degree of fermionic attributes
and  characteristics (shape,size) of the confining potential,
may provide clues  to  the actual contribution of quantum information and
entropy measures to their overall binding. 
The role of quantum information exchanges in complex
systems in the creation of  new forms elementary excitations 
remains unsolved, and is worthy of further investigations.
As far as we are aware, all  known works 
on  the  composite boson system, such as that represented by 
multi-excitonic systems, have excluded the binding forces that may arise
from the energetic value of information. 
The obvious reason for this being that  any information exchanges
 arising from Pauli-based interactions have been
considered ``non-physical", and consequently  the associated retrieval
of   an energy-like quantity from specific interactions have not been given much attention.

\begin{figure}[htp]
    \subfigure{\label{fige}\includegraphics[width=8.25cm]{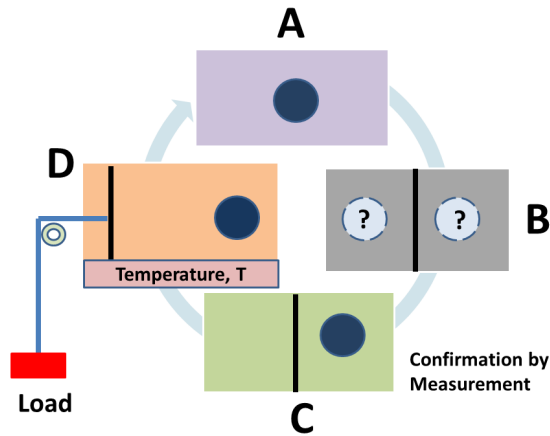}}\vspace{-1.1mm} \hspace{1.1mm}
     \caption{(Colour online) Simplified layout of the thermodynamics of the Quantum Szilard Engine
with 4 main constituents: (A)  Single atom is placed
in an isolated box, (B)  Partitioning into two disconnected regions using the
solid wall. The uncertainty in actual position of the atom is
denoted by ``?", (C) Confirmation of actual position of atom via  measurement 
and lastly, (D) Extraction of  work by
isothermal expansion at $T$ using load apparatus and removal of partition  (when shifted
to edge of box).
 }
 \label{eng}
\end{figure}

\section{Binding energy of  two composite bosons due to the Pauli exchange interactions}

In order to demonstrate the evaluation of binding energy of  two composite bosons,
we revisit the recent work of Kim et. al. \cite{Wook} in which 
the quantum version of the  Szilard engine was considered. 
Kim et. al. \cite{Wook} showed that the crossover from
indistinguishability (true bosonic) to distinguishability (which we 
interprete as loss in bosonic quality) occurs as the temperature increases. 
We thus consider that  bosons assume fermionic features with change in environmental conditions.
We first examine results of the  toy model of a system of  two bosons
confined in a symmetric potential box  of size $L$, with a wall inserted at $x$=$L/2$
at the start.  The  quantum  Szilard engine is based on the  relation \cite{Wook}
 \be
\label{work}
W_{\rm tot}=-k_BT\sum^{N}_{m=0}f_m \; \mathrm{ln} \left(\frac{f_m}{f^*_m}\right)
 \ee
where $W_{\rm tot}$ is the total work arising from the 
wall insertion, measurement and adiabatic wall movements
(Fig. \ref{eng}) performed
by the engine during a single cycle. 
$k_B$ is the Boltzmann constant, $T$ is the temperature of the heat bath and 
$f_m$ is the probability of measuring $m$ particles to the left of the partition.
The term  $f_m^*\coloneqq Z_m(l^m_{eq})/Z(l^m_{eq})$ 
where  $Z_m(l^m_{eq})$ is the partition function of the system with
 $m$ particles on the left,  at the equilibrium position denoted by  $l^m_{eq}$.
This equilibrium position is determined by a delicate balance of forces on the partition \cite{Wook}:
\bea
\nonumber
{\rm Force (left)} &+& {\rm Force (right)}= 0 \\
{\rm Force} &=& \sum_n P_n (\frac{\delta E_n}{\delta X})
\label{bal}
\eea
where  $X$ is an external parameter and $P_n$ is the mean occupation number of
the $n$th eigenstate with energy  $E_n$. It is 
assumed that the system is in thermal equilibrium throughout the engine operation,
and includes the possibility that the particles may in tunnel into the other side of the wall.
For the one particle  ($N$=1) case, the
total work performed by the quantum  Szilard engine during one cycle is
 obtained using  $f_0$=$f_1$=$\frac{1}{2}$, $f^*_0$=$f^*_1$=1
and Eq.  \ref{work} as  $W_{\rm tot}$ = $k_B T$ ln2.

For the two indistinguishable bosons that can be present at
 two possible locations of the box,
using $f_0$=$f_2$, $f^*_0$=$f^*_2$=1 (when the partition is 
pushed to the end), $f_1$=$f^*_1$, the total  work that
can be extracted was obtained as \cite{Wook}, 
\bea
\label{woo}
W_{\rm tot} &=& - 2k_{\rm B}T f_0 \ln f_0,
\\
f_0 &=& \frac{1 + p}{4 + 2 p},
\label{pur}
\eea
where  $p$ is a function of the  temperature, $T$. Here however, we consider the decrease
in $p$ (with increase in $T$) as loss in quality of the bosonic character linked
to ``degree of indistinguishability". 
We note  the importance role of  temperature which is associated with
 thermal fluctuations: it determines whether a boson particle 
acquires  distinguishable features, hence with the availability of a range of electronic
states at higher temperatures, bosons become
increasingly fermionic  and distinguishable. The importance of the  parameter provided
by the approximate separation between energy levels, $\Delta \approx k_B \; T_c$
should be noted. The high temperature regime refers to $T \gg T_c$, while the reverse
holds valid for low temperatures.

 The acquisition of fermionic features with increase in temperature
appears similar  in context to the loss in ``purity" used in earlier
works on composite bosons \cite{Law,Woot,Rama,Tichy1,Tichy2,bobby,thilamc}. 
A system of bosons which are indistinguishable
possess a large Schmidt number ${\kappa}$, and hence can be 
considered  a  highly correlated entangled system. Any deviations
from this arrangement due to changes in the external environment,
leads to irretrievable loss in bosonic indistinguishability.
One can  appreciate that an increasing temperature also has the same
effect as increased confinement of a system of ideal bosons. For
each boson  that constitutes two fermions, as in the case of the exciton,
a display of  the crossover from indistinguishability to
distinguishability for bosons  with change in external parameters (temperature,
confinement) becomes evident.  We therefore quantify  the quality   of the bosonic indistinguishability
using $p$ as it appears in Eq.  \ref{pur}. At  $p$=1, the true bosonic state yields $W^b_{\rm tot} = (2/3)k_{\rm B}T \ln 3$, and at the extreme limit of  $p$=0, where the boson exhibits fermionic features,
$W^f_{\rm tot} = k_{\rm B}T \ln 2$.

In the low-temperature limit, the ground state is predominantly occupied
as higher excitation levels are suppressed. The possibility
that particles tunnels from high to low energy levels
as the barrier is shifted, during the operation of the Szilard engine, is minimized at low temperatures.
Such tunneling effects are expected to dominate at higher temperatures. The importance of including
 the work gained during the removal of the barrier, during which tunneling 
occurs in the event of non-equilibrium partition position, was 
discussed in Ref.\cite{ved1}. 
The same authors  highlighted 
that work extracted is based mainly on  the information gain of the initial measurement \cite{ved2},
which ultimately, is  dependent on the nature (e.g. distinguishable or indistinguishable partition,  boson or fermion) 
of the system being examined. 

\begin{figure}[htp]
    \subfigure{\label{fer1}\includegraphics[width=2.05cm]{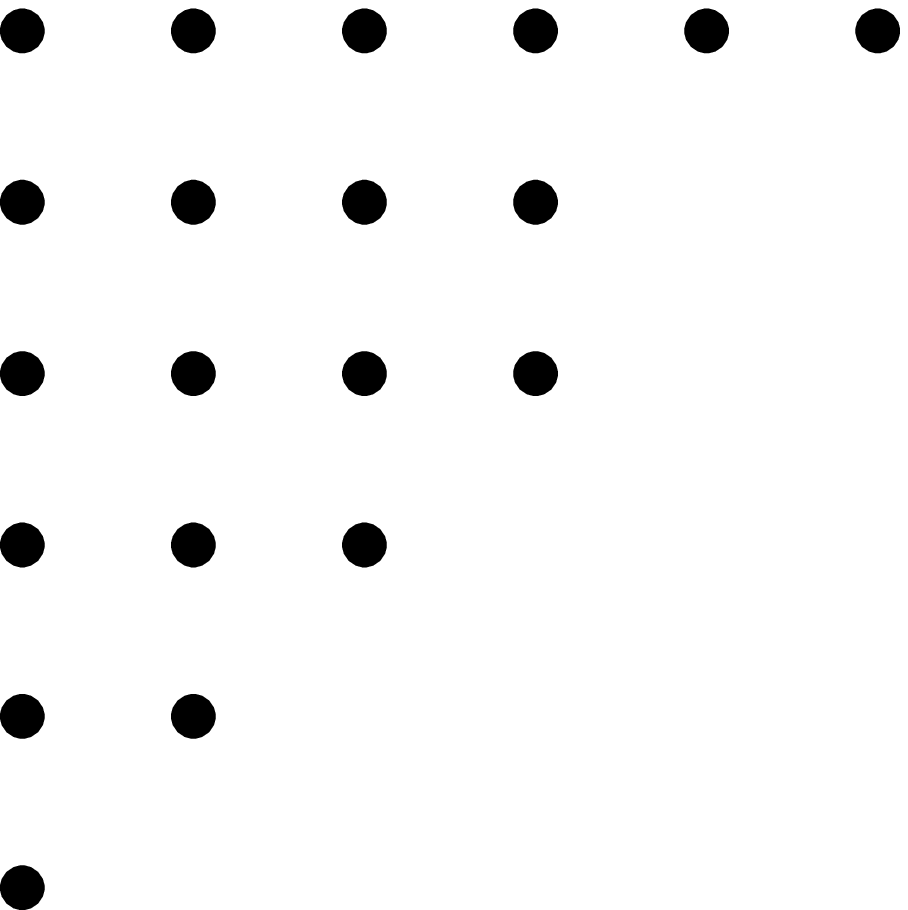}}\vspace{2.1mm} \hspace{4.1mm}
     \subfigure{\label{fer2}\includegraphics[width=2.05cm]{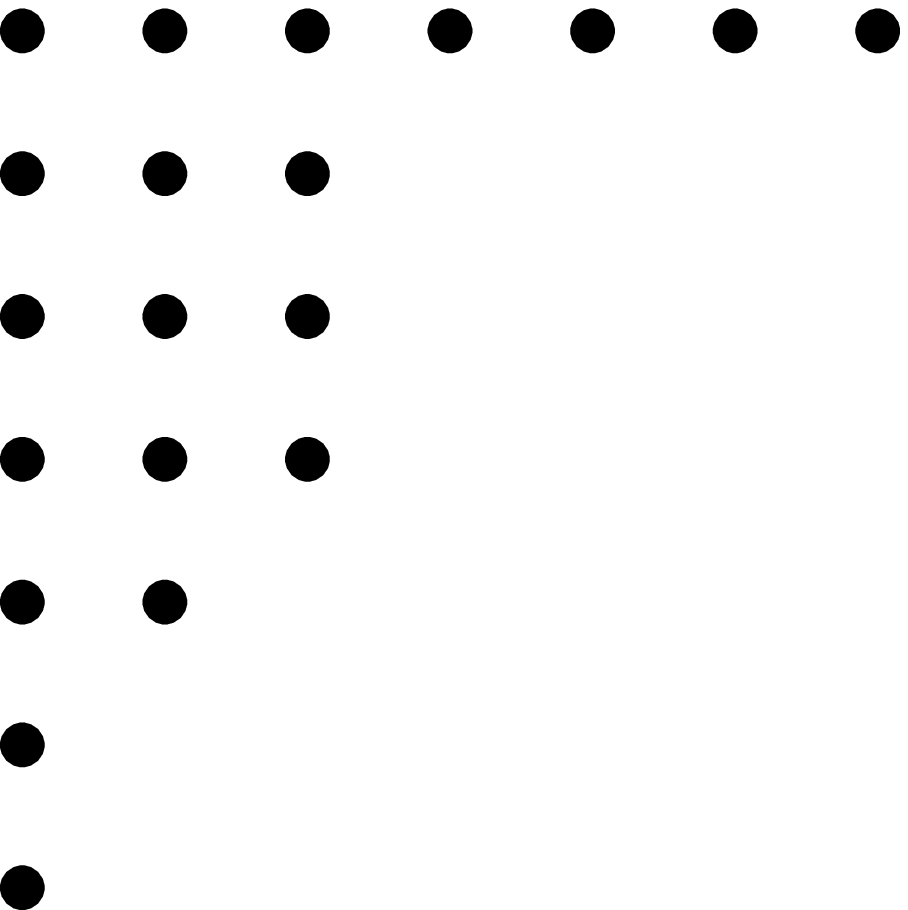}}\vspace{2.1mm} \hspace{4.1mm}
  \subfigure{\label{fer3}\includegraphics[width=2.05cm]{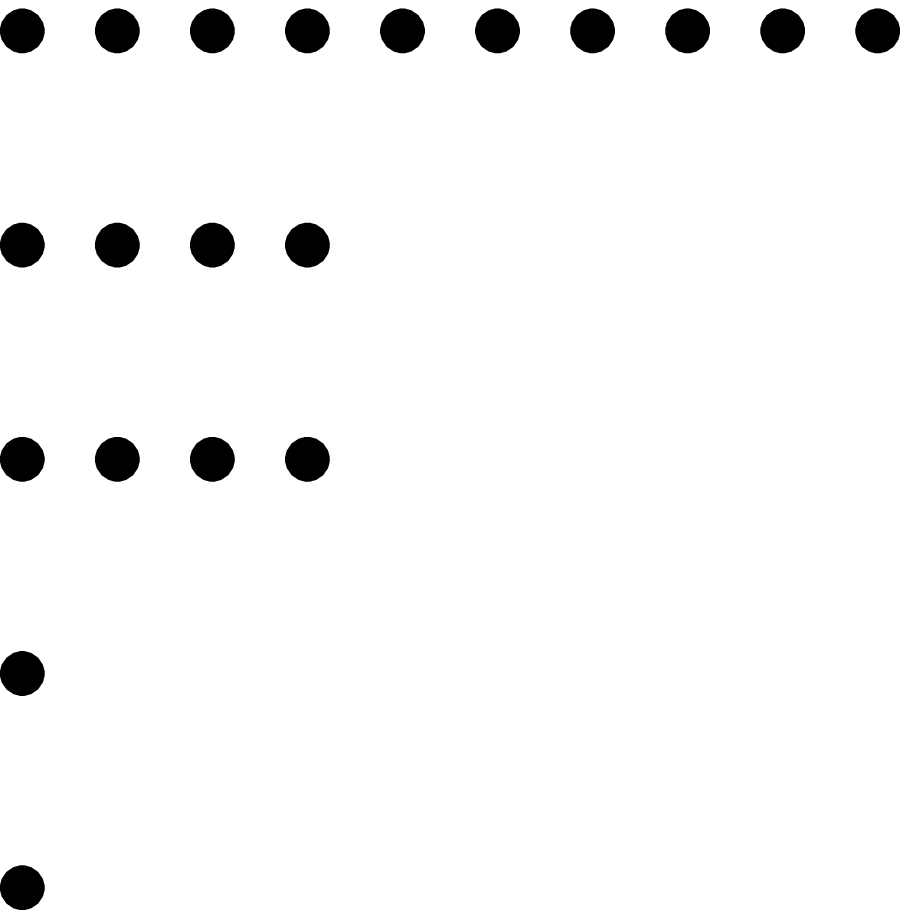}}\vspace{2.1mm} \hspace{4.1mm}
     \subfigure{\label{fer4}\includegraphics[width=2.05cm]{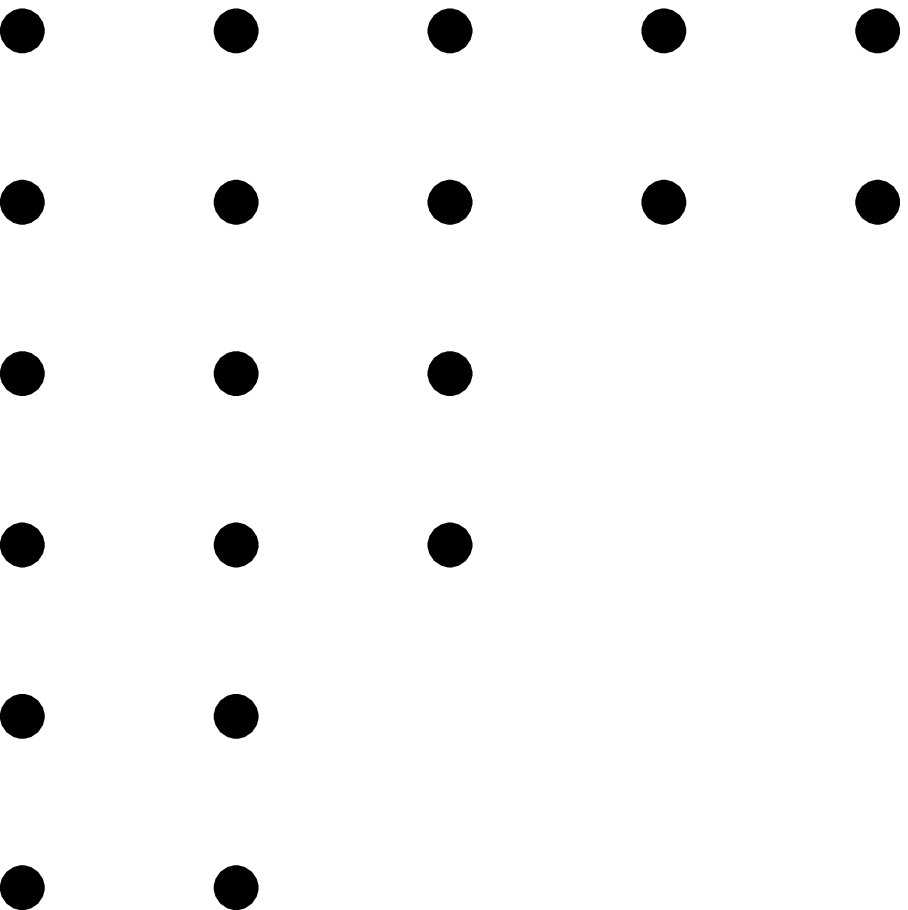}}\vspace{2.1mm} \hspace{4.1mm}
  \subfigure{\label{fer5}\includegraphics[width=2.05cm]{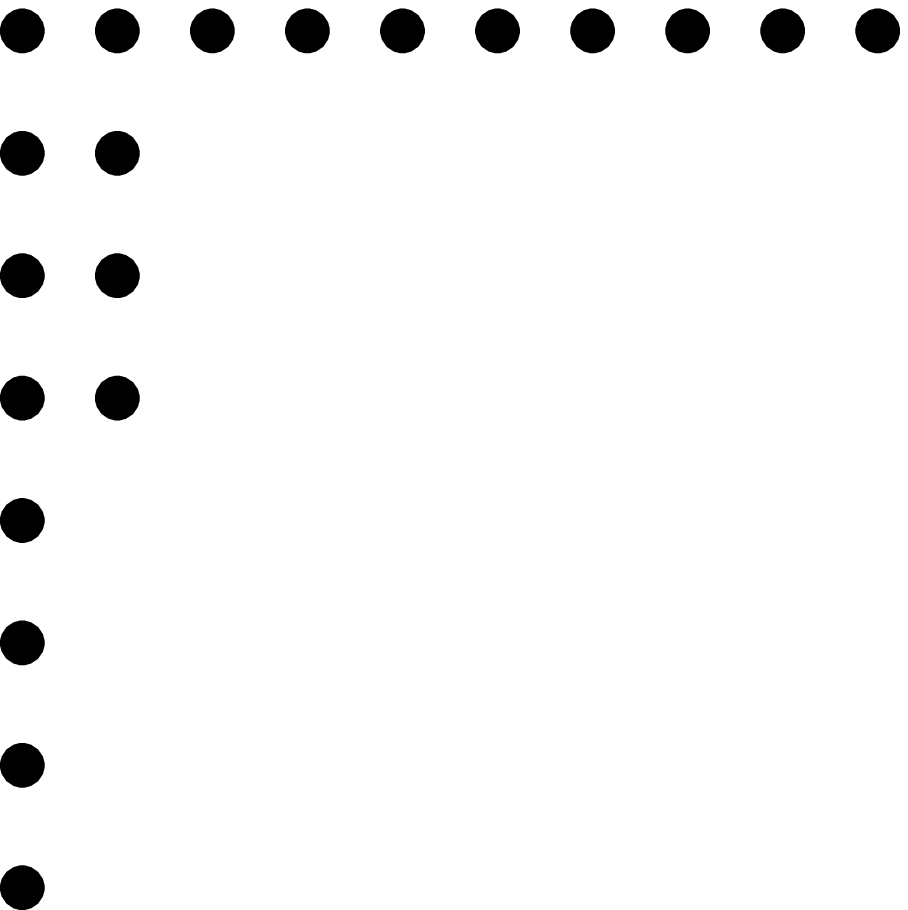}}\vspace{2.1mm} \hspace{4.1mm}
     \subfigure{\label{fer6}\includegraphics[width=2.05cm]{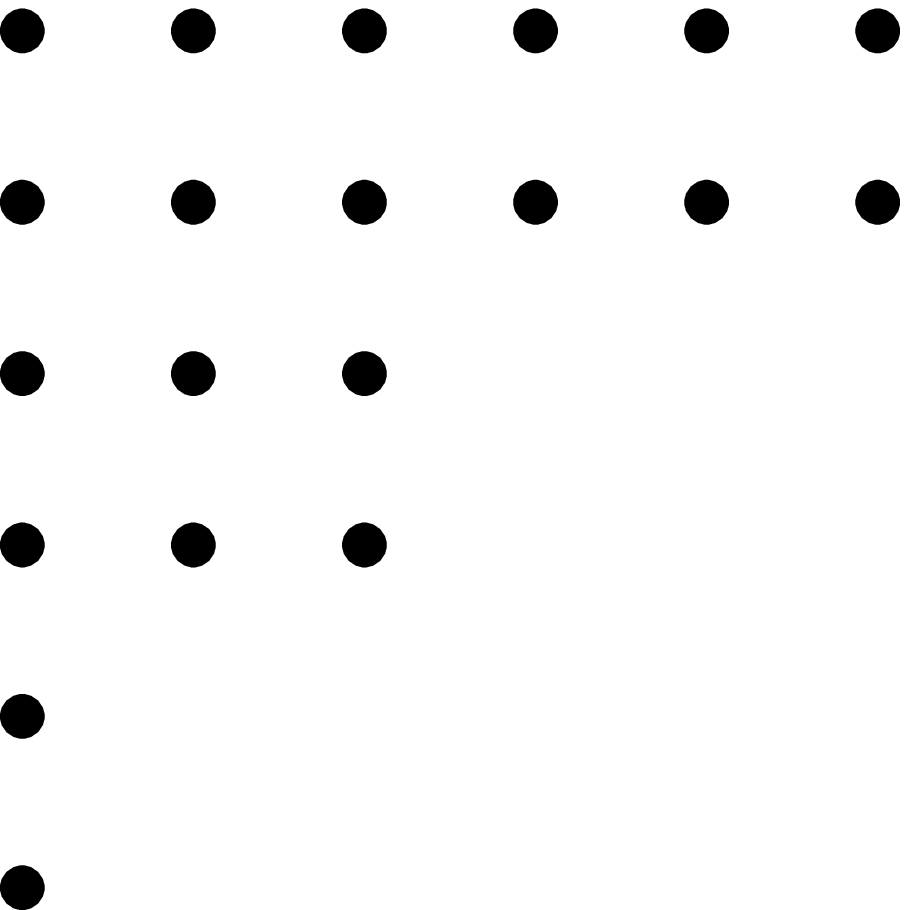}} \vspace{2.1mm} \hspace{4.1mm}
   \caption{Representative arrangement of 20 indistinguishable particles using the Ferrers diagram, where particles
can choose to occupy different energy levels. Each shape corresponds to a specific partition, such as,
(6,4,4,3,2,1), (7,3,3,3,2,1,1), (10,4,4,1,1) etc, with only six possible partitions  shown.
 At the low temperature limit, there occurs  convergence to a specific
partition where all particles occupy one level state, reducing the need to 
incorporate the  tunneling energetics in the Szilard engine. }
 \label{fers}
\end{figure}

Fig. \ref{fers} shows the arrangement of 20 indistinguishable particles
using the Ferrers diagram, which depicts the array of  left-justified solid
circles (e.g, bosons) for a number of partition possibilities.
As the temperature approaches zero,  the possible arrangements converge to one partition
where all particles occupy one level state. The incorporation of  entropy changes associated with tunneling effects based
on the Ferrers diagram (see Fig. \ref{fers})
which shows the different occupation possibilities, is therefore appropriate at higher temperatures,
in more realistic treatments of the Szilard engines.

Bearing in mind that spin attributes were neglected in the original formulation \cite{Wook}
 of Eq.  \ref{work}, we define the composite boson binding energy due to Pauli interactions as
\be
\label{bindp}
E_b(p) = W^b_{\rm tot}(p) - W^f_{\rm tot}
\ee
where the superscript ``b" correspond to a particle with a higher measure of bosonic
quality and superscript ``f" correspond to one that exhibits greater fermionic feature.
This defined binding energy removes, though not entirely,
the tunneling effects (discussed in Ref.\cite{ved1}) that is  common to the terms on the right hand side
of Eq. \ref{bindp}. We note that in excitonic systems, the  binding energy is the 
difference between the coboson energies  and the optical gap \cite{thilex,jai1}.
Fig. \ref{pur} shows the monotonic increase of the binding energy of the boson
with increase in the bosonic quality, $p$. As is well known, 
 more work can be extracted from the ``information-rich"
boson which has the highest quality measure  of unity.

\begin{figure}[htp]
    \subfigure{\label{cap}\includegraphics[width=6cm]{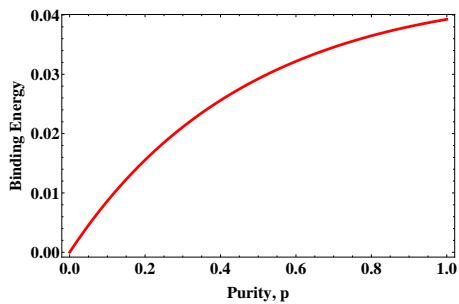}}\vspace{2.1mm} \hspace{4.1mm}
  \caption{Binding energy (in units of $k_B T$) as a function of bosonic quality, $p$ , computed using 
Eqs.\ref{woo} and  \ref{bindp}.
}
 \label{pur}
\end{figure}

\section{The $N$ distinguishable particle systems}\label{Nbo}

\begin{figure}[htp]
    \subfigure{\label{fig3a}\includegraphics[width=3.25cm]{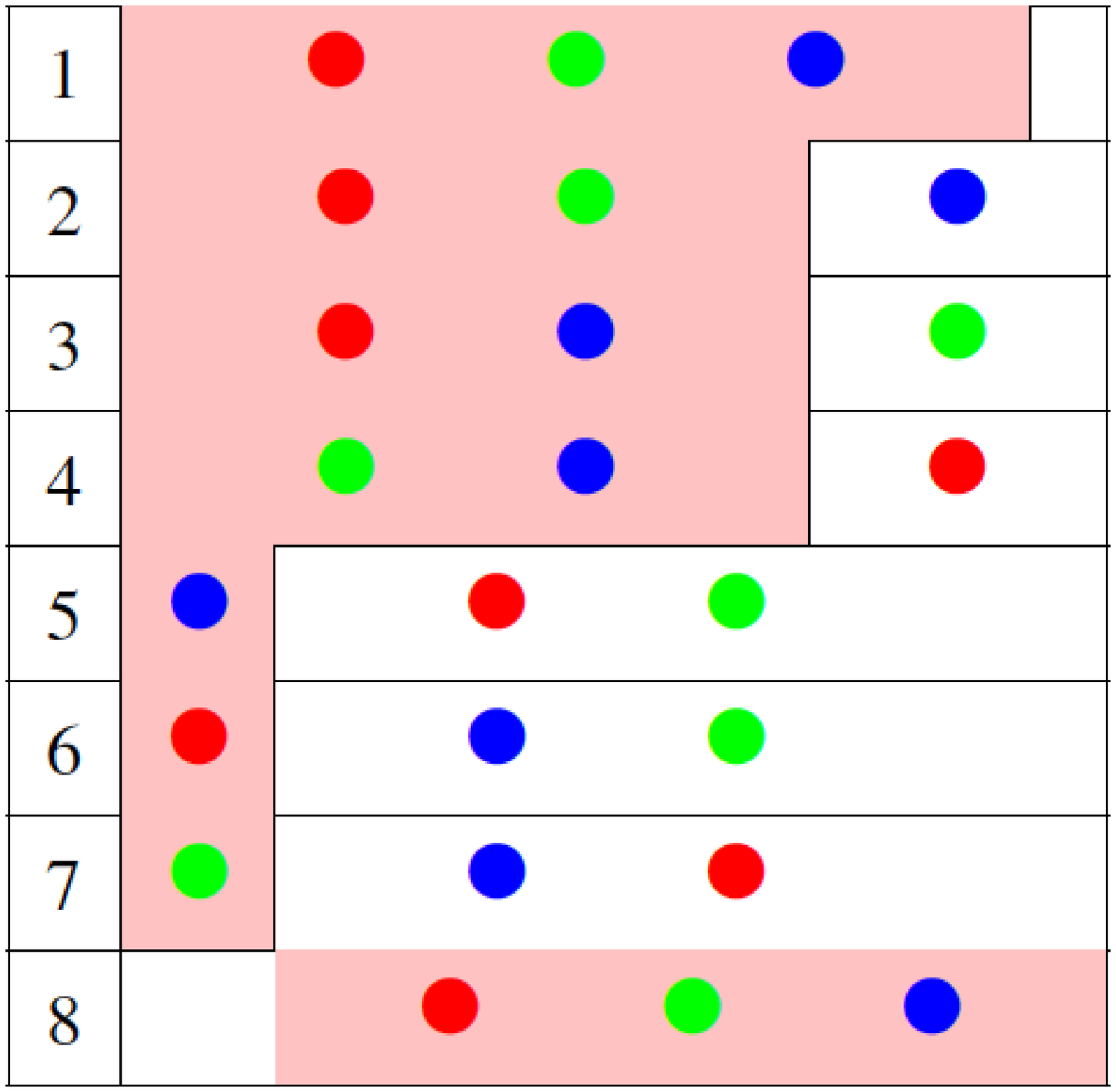}}\vspace{-1.1mm} \hspace{14.1mm}
     \subfigure{\label{fig3e}\includegraphics[width=3.25cm]{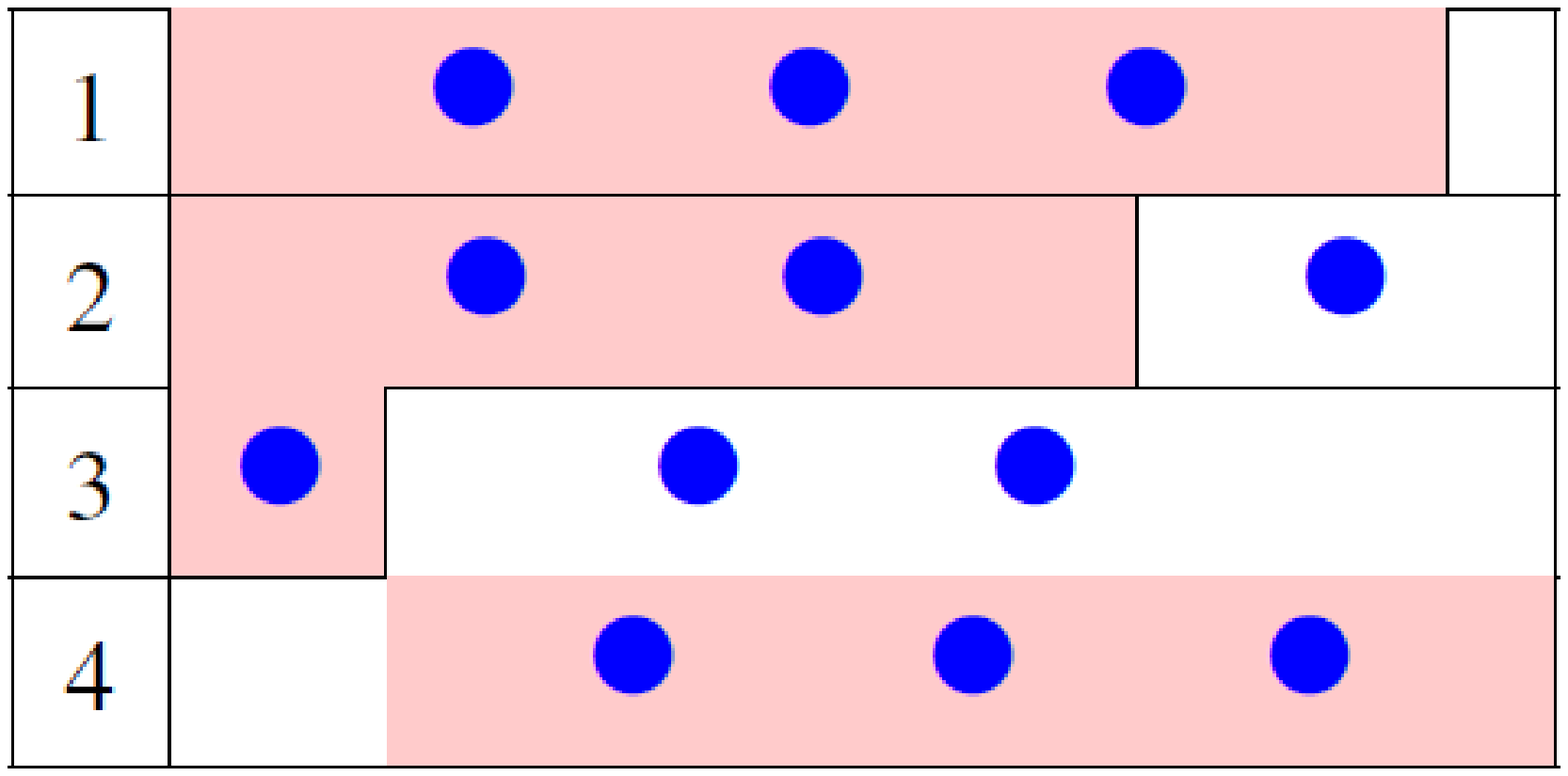}}\vspace{-1.1mm} \hspace{1.1mm}

  \caption{(Colour online) Number of ways that three
distinguishable  (and indistinguishable) particles can be placed within two distinguishable partitioned spaces.
The three distinguishable particles are coded using different coloured balls (red, green, blue).
 }
 \label{box}
\end{figure}

The case of the general $N$ distinguishable particles  enclosed in the Szilard engine, has been examined in 
earlier works \cite{Yao,ved2}, however we reiterate some salient features of generalized systems for the purpose
of examining the binding and capacitive energies, which remains largely unexplored. 
For instructive purposes, we briefly examine the case of three particles in the Szilard engine.
In Fig. \ref{box}, we show the number of ways that three
distinguishable  (and indistinguishable) particles can be placed within two distinguishable partitioned spaces.
For the general $N$ particles, the total number of microstates is $\Omega_N$=$2^N$ since each particle can occupy one of two possible
locations,  we thus obtain  $\Omega_3$=8. For $m$ probable number of particles on the left partition, we obtain
the total multiplicity, 
\be
\Omega_N^m= \frac{N!}{m!(N-m)!}
\label{lep}
\ee
A general expression
for $f_m$ is  given by
\be
f_m = \frac{\Omega_N^m}{\Omega_N} = \frac{N!}{2^N m!(N-m)!}
\label{fm}
\ee

At the equilibrium point, the total number of microstates is determined  by
the probability $m/N$ of particles on the left and $(1-m/N)$ on the right side
of the engine, hence
\be
f^*_m =  \frac{N!}{ m!(N-m)!}\left(\frac{m}{N} \right)^m  \left(1-\frac{m}{N} \right)^{N-m}
\label{fm2}
\ee
 $f_m$ (Eq.  \ref{fm}) and $f^*_m$ (Eq.  \ref{fm2}) applies to boson and fermions
 in the high temperature limit, due to increased availability of extra energy states.
In the case of three distinguishable particles, we obtain $f_0$= $f_3$=$\frac{1}{8}$, $f_1$=$f_2$=$\frac{3}{8}$.
At equilibrium, it is easily  shown that $f^*_0$=$f^*_3$=1, while at high temperatures, the value of $f^*_1$=$f^*_2$=$\frac{4}{9}$ can be obtained using Eq.  \ref{bal}. 

By setting $m'=N/2-m$, and using Eq.  \ref{lep} and the Stirling formula $\mathrm{ln}\; m! \approx m \; \mathrm{ln}\; m-m$, we obtain
for a small $m'$, $\mathrm{ln}(\Omega_N^{m'}) \approx -\frac{2 {m'}^2}{N} +N \mathrm{ln}\; 2$ or
\be
f_{m'} = {\rm exp}(-\frac{N}{2} \alpha^2)
\label{fmpr}
\ee
where $\alpha$=$\frac{2 m'}{N}$. Hence the distribution, $f_m$ becomes 
Gaussian in the limit of very large $N$, a result that is well known in the standard
statistical physics text \cite{sbook}. Eq.  \ref{fmpr} indicates
that  it is more likely that the particles are 
equally distributed on either side of the 
box,  (i.e. $m$=$\frac{N}{2}$ correspond to the most probable microstate), and $f_{\frac{N}{2}} \rightarrow 1$. 
At equilibrium position, $f^*_{\frac{N}{2}} \rightarrow 1$ as well, thus the total work that can be extracted
at  $N \rightarrow \infty$ is 
\be
W_{\rm tot}=  0
\label{wze}
\ee
where we neglect the small population difference in the case of an odd number of particles.
 Eq.  \ref{wze} is the result of  minimal changes in population (and information) imbalances
at start and end of the engine cycle, and also due to the choice of an  initial measurement
in which the partition is inserted in the middle of the Szilard engine.
Thus far in  the analysis involving various entropy changes,
we have employed a simplified model where information content associated with
tunneling of the particles within the various energy levels (see Ferrers diagram
in Fig. \ref{fers}) was neglected.

\subsection{Biased Szilard machine}

Here we consider the  existence of  a biased distribution of quantum particles
at the beginning of the cycle. This may occur as a result of a $r \%$  preference for particles to be
located in the left side of the engine, as compared to the right side, due to variations in 
available energy levels. 
We thus rewrite Eq.  \ref{fm} as
\be
f^{bi}_m = \frac{\Omega_N^m}{\Omega_N} = \frac{N!}{ m!(N-m)!} \left(\frac{1}{2}+r \right )^m \left(\frac{1}{2}-r \right )^{N-m}
\label{fm}
\ee
where the superscript $bi$ denotes a biased Szilard machine. Assuming the presence of large number
of participants, $N$ and a equilibrium state where the biased arrangement is absent,
we obtain, using the  Stirling formula and Eq.  \ref{work},
 \be
\label{workb}
W_{\rm tot}=-k_B \;T \; \mathrm{ln} \left(1- \beta^2 \right)
 \ee
where $\beta$ = $N r \ll 1$, and  we set $f^*_{\frac{N}{2}} \rightarrow 1$ at the  equilibrium point of the engine.
The non-zero work extracted here arises due to the biased distribution at the start of the
cycle, and is a result of  energy that is transferred from the 
``observation energy" that was expanded in order to introduce a  preferred distribution of particles.

\subsection{Szilard engine with indistinguishable partitions}
In the case of a quantum szilard machine, where the partitions are indistinguishable,
i.e, we are unable to tell the left box from the right box, the total
number of possible microstates is decreased further. 
For instance, the possibilities of arrangements numbered  1 and 8 in 
Fig. \ref{box} become equivalent (or $f_0$= $f_3$=$\frac{1}{4}$),
as the total number of partitions is 4.
Likewise those numbered 2 to 4 
can be merged with options 5 to 7, and we get $f_1$= $f_2$=$\frac{3}{4}$.
  For the general case, where   $N$ distinguishable balls
distributed into the 2 indistinguishable sides of the Szilard engine, the 
total number of distribution possibilities  is given by
\be
\label{ind}
\Omega_N= 1+ S(N,2)
\ee
where  $S(n,m)$ is the  Stirling numbers of the second kind \cite{stir1,stir2}.

\section{$N$ indistinguishable particle systems: Information Capacitive  Energies}\label{cap}
Unlike in  Section \ref{Nbo} where we considered distinguishable particle systems, 
 here we obtain some estimates of the information
capacitive (or addition) energies of $N$ indistinguishable particles  such as bosons.
Following the  relations for capacitive energies
in Ref.\cite{ind}, we write 
\bea
\label{cap1}
E^1_c(N) &=& W^b_{\rm tot}(N+1) - W^b_{\rm tot}(N)\\
\label{cap2}
E^2_c(N) &=& W^b_{\rm tot}(N+1) + W^b_{\rm tot}(N-1)-2 W^b_{\rm tot}(N)
\eea
$W^b_{\rm tot}(N)$,  the work extracted from a system of $N$ bosons,
at low temperatures, is obtained by
considering the case of $N$ indistinguishable particles placed in two distinguishable
partitions where  the total   
number of microstates is given by $\Omega_N={N+1}$, with
\be
f_m (T \rightarrow 0) = \frac{1}{N+1}
\label{fm}
\ee
As shown in Fig. \ref{box}b, the total number of possible arrangements in the case of $N$=3,
 is given by $\Omega_3$=4,
with $f_0$=$f_1$= $f_2$=$f_3$=$\frac{1}{4}$.
$f^*_m$ is dependent on the shape of the confining potential,
and has been evaluated for the infinite potential well as \cite{Yao}
\be
f^*_m (T \rightarrow 0) = \exp \left[-\frac{m}{k_B T} \left(E_0(l_e^m)-E_0(L-l_e^m) \right) \right ]
\label{fmY}
\ee
for $0 < m < N-m$, and  where the initial 
location  of the partition is set at $\frac{L}{2}$.
It is obvious that
 $f^*_{\frac{N}{2}}= \frac{1}{N+1}$, $f^*_{0}$=$f^*_{1}$=1.

We consider the   one-dimensional cluster of $N$ bosons trapped  in the
potential that is approximated by the harmonic oscillator well, noting that
aside its relevance to  fundamental studies, one-dimensional systems can be fabricated
 by restricting the boson degrees of freedom in the  transverse or radial directions.
The radial confinement energy  then becomes dominant
compared to other system energies such as $k_B T$ and 
 the axial confinement trapping energy $\hbar \omega$. Experimental
state of condensates (of about 10$^4$ atoms)  have been realized using  elongated magnetic
traps \cite{exptharm} with respective radial and axial trapping
frequencies of 360 Hz and 3.5 Hz.

In the harmonic oscillator well, 
trap length of the Szilard machine equals 
\be
L_t =\sqrt{\frac {\hbar}{m_b \; \omega}}
\label{trap}
\ee 
where $m_b$ is the 
mass of bosons and  $\omega$ is the axial confinement frequency.
Based on  the convergence of all possible arrangements of particles
in different energy levels to one partition
of  particle occupation (example of 20 particles shown in Ferrers diagram of Fig. \ref{fers}),
we obtain using Eq. \ref{bal}, a
simple expression for the equilibrium length, $l_e^m$
\be
\label{leneq}
l_e^m=\frac{L}{1+ \left( \frac{N-m}{m} \right)^{\frac{1}{3}}}
\ee 
Using Eqs.\ref{work}, \ref{fm} and \ref{leneq}, and an analogous expression
to  $f^*_m$ (Eq. \ref{fmY}) for the harmonic oscillator well,
we obtain (in units of $k_B \; T$) the total work that can be extracted from
a condensed state of $N$ bosons  at very low temperatures as
\be
\label{wkbo}
W^b_{\rm tot}(N) = \alpha \; {\rm ln}(N+1) - \frac{2 \gamma}{N+1} \sum_{m=1}^{\frac{N}{2}-J} m \left[
\left(1+ \left( \frac{N-m}{m} \right)^{\frac{1}{3}} \right)^2-
\left(1- \frac{1}{1+ \left( \frac{N-m}{m} \right)^{\frac{1}{3}}} \right)^{-2} \right ]
\ee
where $\alpha$ = 1 ($\frac{N}{N+1}$) for odd (even) number of bosons,
and $J$=$\frac{1}{2}$ (1) for odd (even) number of bosons. 
The dimensionless quantity, $\gamma \approx \frac{\hbar \omega}{k_B \; T}$, hence 
 Szilard engines with  small trap lengths give rise to higher confinement energies,
resulting in larger values of $\gamma$. It is obvious that for weakly confined particles 
(large trap lengths), the second term in Eq.  \ref{wkbo} can be neglected, and the
total work that can be extracted increases monotonically with $N$.
The basis for the slight differences
between the odd and even number of particles is due to the asymmetrical  partitioning
in the vicinity of $\frac{N-1}{2}$ for the odd number of bosons. This attribute has been 
examined as a parity effect in  a recent work by Lu et. al. \cite{Yao}  for 
 the $N$-particle quantum Szilard engine, for bosons and fermions, using the infinite potential well.
As noted earlier in Eq.  \ref{wze}, the difference due to  the odd and even number of particles
in the Szilard engine become negligible in the limit of very  large $N$.

\begin{figure}[htp]
    \subfigure{\label{cap}\includegraphics[width=7cm]{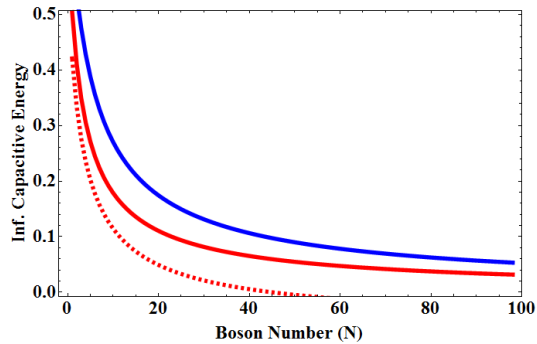}}\vspace{2.1mm} \hspace{4.1mm}
  \caption{Information capacitive  energies at low temperatures (in units of $k_B T$), computed using 
Eqs.\ref{cap1}, \ref{cap2} and  Eq.  \ref{wkbo} as function of boson number $N$.
Blue solid line correspond to energies computed using Eq.  \ref{cap2}, while
red solid line ($\gamma$=0) and red dashed line ($\gamma$=0.1)
correspond to energies computed using Eq.  \ref{cap1}. The dashed lines
highlight the influence of confinement effects on binding energies}
 \label{capa}
\end{figure}

In Fig.\ref{capa}, we plot the information capacitive energies using  Eqs.\ref{cap1}, \ref{cap2}
and $W^b_{\rm tot}(N)$ in Eq.  \ref{wkbo} as function of boson 
number $N$. There is  gradual decline of the capacitive energy
with $N$, which  indicates that the information content  associated with
an additional boson diminishes with the size of the boson cluster, as
quantified by $N$.  Hence the work that can be extracted from an extra boson decreases with 
the number of bosons already contained within the Szilard engine.
 In the presence of confinement effects 
(non-zero $\gamma$ in Eq. \ref{wkbo}), the extractable work vanishes at a critical $N_\gamma$,
due to boson particles acquiring fermionic features . Hence the confinement
effects introduces a limit to the size of the total number of boson 
in the Szilard engine, from which work can be extracted. It is to be 
noted that these results are obtained in the low temperature regime,
and are therefore not representative of trends at high temperatures. 

We note that the defined information energies in Eqs\ref{cap1}, \ref{cap2}
are not unique, as analogous definitions can be obtained for 
systems that consist of $N_b$  bosonic particles and $N_f$
 fermions. If $N_f$ is even, then total energy that
can be extracted depends,  solely on $N_b$ bosons
within the confines of  simplifying assumptions at low temperatures.
In the case where $N_f$ is odd, 
 an additional energy, $K_B$T, will need to be accounted for,
this will be discussed in the  next Section.

\section{Confinement effects on the binding energies of $N$ bosons}\label{conf}
Following  the results presented in Section \ref{Nbo}, we now examine the binding energies
of a system of boson at low temperature by extending Eq.  \ref{bindp} to  an
$N$-particle system
\be
\label{binds}
E_b(N) = W^b_{\rm tot}(N) - W^f_{\rm tot}(N)
\ee
where  $W^b_{\rm tot}(N)$ appears in Eq.  \ref{wkbo}, and 
only  $W^f_{\rm tot}(N)$ needs to be obtained to evaluate the
binding energy, $E_b(N)$.
As in the case of $W^b_{\rm tot}(N)$ for  bosons (see Eq.  \ref{wkbo}),
the extractable work, $W^f_{\rm tot}(N)$   depends on whether it is an odd
or even number of fermions present in the Szilard engine.
Due to prohibition of occupation of the same state by two fermions by virtue of the
exclusion principle, 
even number of fermions are more likely to be distributed equally  over the two sides of the 
partition.  In this case,
no work can be extracted from an even number of fermions at 
low temperatures, as   isothermal expansion does not occur with the same number of  particles  on either  side of the wall.
As  is well known, and also indicated by  Eq.  \ref{wkbo},
bosons are not subject to these same rules. 

\begin{figure}[htp]
    \subfigure{\label{bi}\includegraphics[width=7cm]{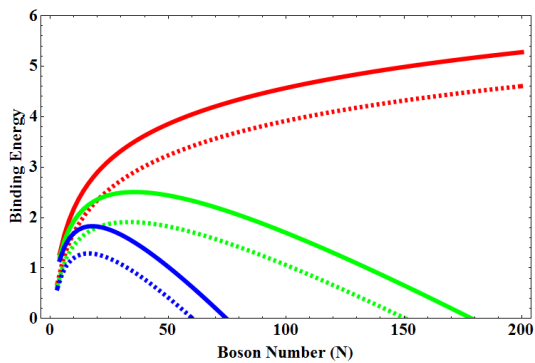}}\vspace{2.1mm} \hspace{4.1mm}
  \caption{Bosonic binding energies (in units of $k_B T$) at low temperatures, 
computed using Eqs.\ref{binds}, \ref{wkbo} and \ref{workodd} as a function of $N$,
 for various values of the confinement parameter, $\gamma$
(top red (0), middle green (0.05), and bottom blue (0.1)). The solid lines correspond to 
even number of bosons, while dashed lines correspond to odd number of bosons. }
 \label{bindN}
\end{figure}

The low temperature
characteristics of fermions have 
 been highlighted by Kim et. al. \cite{kim2}
using the third law of thermodynamics,
which predicts that the entropy becomes zero
in the absence of degeneracy in the ground state.
Otherwise a non-zero residual entropy is retained,
as is the case for the odd number of fermions.
In the latter case, 
there exists  some degree of uncertainty with respect
to the  unpaired fermion, and  there appears ground state
degeneracies  at low temperatures. The work performed during a single engine cycle 
for the odd number of fermions is independent of  $N$
\be
\label{workodd}
 W^f_{\rm tot}=-k_BT \mathrm{ln} \left(\frac{1}{2}\right)
= k_B T \; \mathrm{ln} 2
 \ee
In Fig.\ref{bindN}, we plot the binding energies computed using Eqs.\ref{binds}, \ref{wkbo}
and \ref{workodd} as a function of $N$ for various values of the confinement parameter, $\gamma$
(see Eq. \ref{wkbo}). The difference in the binding energies due to the odd and even number
of particles, is  attributed to the residual entropy retained by odd number
of fermions, as discussed earlier. An increasing confinement effect obviously results
in a decrease of the binding energy of the system of  bosons, as it acquires greater
fermionic characteristic features. In other words, increasing confinement effects
results in decrease in the bosonic quality (similar to the decrease in the measure $p$ in 
Eq.\ref{pur})

The results obtained in Fig. \ref{bindN} may be 
compared to the binding of excitonic systems in quantum dot
structures \cite{triex}. Recently, several sharp peaks, indicating the
 multi-excitonic complexes (including the charged triexciton)
 have been  observed in in the photoluminescence
spectra of self-assembled
GaAs quantum dot systems \cite{ara}. The triexciton,
consisting  of three electrons and four holes, as well
as singlet and triplet states of charged biexcitons  have been  identified in the photoluminescence
spectra \cite{ara}, showing typical exchange splittings
of 120 $\mu$eV (0.12 meV). 
The micro-photoluminescence procedure  was performed on  quantum dots with
base size of 30 nm, which were kept at 8 K in a
cryostat. If we assume each quantum dot to be a miniature Szilard engine, 
and taking $N \approx$ 2 to 5 (which gives binding energy of order of $k_B T$ in 
Fig. \ref{bindN}), one obtains $k_B T \approx 0.7$ meV at 10 K. 
This  energy estimate is obtained by neglecting confinement effects
associated with the base structure, nevertheless it is of the same order
of magnitude as the experimentally observed exchange splitting energies.
This indicates that there may be non-negligible  contributions
 of energy like quantities arising from information exchanges during
quantum measurements. 

Based on the results obtained in Section \ref{conf}, we note that 
energies of the order of  0.1 meV can be associated with the residual entropy of
odd number of fermions at low enough temperatures. This is about the same order
of  interaction energies noted in  quantum dot excitonic systems \cite{ara}.
Hence, it not unlikely that
exchange interactions of multi-excitonic complexes
may include a non-trivial component  of quantum information theoretic 
entities. To differentiate  various energies involved is however challenging,
in view of the resolution of  existing technologies needed to perform this task.
 To this end, there is need to reexamine the role of the monitoring instruments
during  measurements  of exchange interaction energies, for
accurate interpretation of experimental results. 
Further investigations of  Pauli exchange interactions
of multi-excitonic complexes will contribute to   understanding
of  various recombination mechanisms and the polarization properties of photons,
essential for quantum information and processing applications \cite{niel}.

\section{Bose-Einstein condensate in a Szilard engine}

We explore the feasibility of constructing a quantum Szilard engine
using one-dimensional Bose-Einstein condensates  systems in this Section.  Currently, 
Bose-Einstein condensation is only possible in three-dimensional homogeneous systems, with no definite
 transition noted in unconfined one-dimensional systems \cite{kos}. 
In the case of trapped one-dimensional boson systems however, it was proposed
that   ground state condensates may exist at enough high densities \cite{kette,reca},
although to date, there has been no convincing experimental evidence
in these low-dimensional systems.
Strongly repulsive one-dimensional bosonic systems 
can be mapped to a noninteracting fermionic system.
Hence at lower densities,  a correlated state (Tonks-Girardeau gas) \cite{tonk,lieb} appears
 on the basis of the  Lieb-Liniger  model of impenetrable bosons, and the
repulsive interactions between bosons appear to  simulate the Pauli
exclusion forces between fermions. 
The dependence of the  Bose system on dimensions of the confining potential,
shows the important role  of the  dimensionality of
the Szilard engine. In order  to compare results obtained
in earlier sections, we closely follow the one-dimensional Bose-Einstein condensate
case for evaluation of some energy estimates.

During boson condensation, there is  rapid accumulation of a substantial
 fraction of  particles into the ground state, below a finite critical temperature.
Bose condensation becomes favorable when 
 the density of particles reaches a point when there is at least
one particle per de Broglie wavelength. 
In general, there occurs  insignificant fluctuations about the average 
condensate population at temperature below a finite value. 
A microcanonical approach to the population fluctuations, $\delta N$ of  $N$ ideal
bosons trapped in a one-dimensional harmonic potential shows
that $\delta N \approx \frac{\pi}{\sqrt{6}} \frac{k_B T}{\hbar \omega}$
for  $T \ll \frac{\hbar \omega}{k_B} \frac{N}{ {\rm ln} N}$  \cite{gross1,gross2}.
If we model a Bose condensate system as a quantum Szilard engine,
  fluctuations in particle number will be kept minimal, at low operating temperatures.
In the limit of large N, an expression linking  the number of
atoms and the transition temperature $T_c$ in one-dimensional systems
has been obtained as \cite{kette}
\be
\label{tran}
N = \frac{k_B T_c}{\hbar \omega} {\rm ln}  \frac{2 k_B T_c}{\hbar \omega} 
\ee
Setting the  trap-length, $L_t$, of the 
harmonic well Szilard machine at 100 $\mu$m,
we can evaluate $\hbar \omega$ using  Eq.  \ref{trap}.
Using the typical excitonic polariton mass as
1 $\times 10^{-5} m_0$ where $m_0$ is the free electron mass, 
we obtain using Eq.  \ref{tran}, for given  $N \approx 10 ^3$,
a transition temperature of $T_c$ that occurs at $\approx$ 10K.
The inter-particle distance is computed as $\frac{L_t}{N} \approx$
0.1 $\mu$m.

There is formation of a collective state of Bose-Einstein condensate
when the de Broglie wavelengths of particles given by
\be 
\lambda_{db} = \sqrt{\frac{2 \pi \hbar^2}{m_b k_B T}}
\label{dbro}
\ee
becomes comparable or exceed the inter-particle distance,
$l=\frac{1}{\rho^{1/3}}$, where 
$m_b$ is the mass of the boson, and $\rho$ is the particle number density.
For the excitonic polariton at 10 K, we obtain using Eq. \ref{dbro},
$\lambda_{db} \approx$ 5 $\mu$m at 10K, which increases to about 
17 $\mu$m at 1K.  We thus obtain
an estimate of $\lambda_{db}$, that far exceeds
the inter-particle distance (0.1 $\mu$m), noted earlier
for $10^3$ particles in a Szilard engine of 
trap-length, $L_t$= 100 $\mu$m. These results, even if it is considered
for one set of model parameters,  highlight the favorable conditions
for Bose condensation to occur, especially in highly anisotropic  micro-cavities traps
where the degrees of freedom are essentially erased in two possible 
direction, and in which the polariton species  is introduced.

It is instructive to compare the results obtained here with the 
information capacitive  energies computed (in units of $k_B T$),
at low temperatures, as  shown in Fig. \ref{capa}.
We note that the Szilard engine of trap-length, $L_t$= 100 $\mu$m
would yield the confinement parameter $\gamma$ (see Eq.  \ref{cap2})
values of about 0.04 (at 1 K) and 0.004 (at 10K). These small
confinement measures indicate that the condensate would experience
non-zero capacitive energies of about 0.008 meV, and
binding energies $\approx k_B T$ (Fig. \ref{bindN})
that arise due to  quantum informational entropy
changes. We point out that small trap-lengths will result in confinement
effects reducing binding effects (Fig. \ref{bindN}), thus there is a lower
limit to the size of the length,  $L_z$, if such one dimensional
 Szilard-model systems are to be realized in future experiments.
It is expected that observations can be made
 at the  cryogenic temperatures of a few Kelvins,
which is already accessible with modern technologies.

On the basis of the results obtained so far, 
we may make some pertinent queries, with respect to the model system 
of   polariton condensate in a Szilard engine.
Can quantum informational entropy changes and quantum mechanical effects partly
 (or even substantially) explain the observations of polariton condensates?
It may seem that  energy derived from the Szilard engine via the process of
 observation, may help sustain at least a fraction of the condensation process.
To what degree do such changes, if any, are  introduced
during observations during quantum measurements and contribute to the
Bose-Einstein condensation of material systems? Is there an interference
effect that arises due to the monitoring  apparatus? The answers
to these questions are not immediately
clear and needs further scrutiny on the delicate roles played by
monitoring instruments, and are best answered via direct
experimental verification. 

It is not certain whether quantum
 information that  manifest in the form of entanglement can provide
unification between condensation phenomena and correlation
entities. In a recent work, the generality of 
Landauer's principle to quantum situations was considered \cite{lidia},
in which the role of the observer was emphasized.
In the event that the  observer and system are entangled,
 thermodynamic entities linked  to conditional entropies will arise,
giving rise to a cooling effect on the environment.
When applied to the condensation process taking place within the Szilard engine,
this suggests, that any increased heat generated 
due to ``bunching" of a large number of particles  into a single quantum state,
may be reduced by the presence of an entangled observer. 
To this end, there is 
ample scope to examine the role of irreversibility during Bose condensation,
and to probe  excitations that  arise due to 
the measurement effect in future investigations.

Lastly, we note that unlike the polariton species, the  exciton 
gas whose constituent possess a higher mass,  appears to have
less likelihood of achieving the condensed state
at comparable (few Kelvins)  temperatures. Till now, the exciton Bose-Einstein condensate
has not been directly observed  due to the  critical
conditions for the condensation process \cite{timo}. 
As pointed out in Ref.\cite{bosecom},
excitons exhibit  Bose-Einstein condensation only if these quasiparticles
have long lifetimes, and 
exciton-exciton interactions become repulsive to prevent the
formation of larger exciton systems such as biexcitons.
While the long exciton lifetimes is achievable in highly confined
systems, it is not certain whether biexciton formation can be avoided.
For the latter reason, the formation of a biexciton-polariton quasiparticule
is also likely, and may act as an obstacle to condensation phenomenon
involving the pure (and ideal) exciton gas system. Hence
the conditions which favor the depletion of condensates due to  formation
excitonic complexes have to be eradicated before possible observations of the condensed
phase of exciton gas can be contemplated.

\section{Conclusion}\label{conf}
In summary,  the concept of the Szilard engine which  extracts work 
from a thermal reservoir, via quantum  measurements 
of the particle  position, is used to define the binding energies of composite bosons that
possess some degree of fermionic features. The significance of the quantum Szilard engine is
that it allows the evaluation of 
 binding energies that can be associated with quantum information 
linked to the indistinguishable and/or distinguishable nature of a system of
particles.  The non-trivial role of the quantum information thermodynamic operation
of the Szilard engine during Bose-Einstein condensation of
 the light mass   polariton quasiparticle has been examined, along with
applications related to the 
 binding energies of large multi-excitonic complexes in quantum dot systems.
We allude to the possibility that
quantum informational entropy changes may  contribute to the formation of polariton condensates,
however further advancement in instrumentation techniques, with ability
to   differentiate  different forms of energies, is needed to verify these extensions.

Although in this work,  we
have only focused on two specific systems: Multiexcitonic quasiparticles and 
Bose-Einstein condensates, the developed
methodology is  general and might be 
useful for improving density functional techniques of strategic material systems.
The link between information and energy may  be extended to 
improve  ${\it ab \; \; initio}$ methods  based on  
Kohn-Sham density functional theories (DFT) \cite{kohn,parr,BLYP,PBE}. 
While it is known that density functional theories provide accurate results 
 in systems with weak electronic correlations, many density functionals 
fail to provide adequate description of strongly correlated electronic systems \cite{fail}. 
The use of the quantum Szilard model in the latter
systems  may reveal new forms of
elementary excitations that  arise due to information processing.
These extensions are expected to 
contribute to  novel 
quantum processes that underpins superior  properties  of strongly correlated
systems, with potential applications in optical devices and 
 quantum information processing.

\section{Acknowledgments}

The author  gratefully acknowledges the  support of  the Julian Schwinger Foundation Grant,
JSF-12-06-0000.  The author would like to thank Monique Combescot for earlier discussions
related to the binding energies that arise from Pauli interactions, and 
Malte Tichy for useful correspondences regarding specific properties of composite bosons,
and for pointing out the use of Szilard engine model in Ref.\cite{bobby}, at the time of preparation of 
this manuscript.

\end{document}